\documentclass[preprint]{elsarticle}
\usepackage{lipsum}
\makeatletter
\def\ps@pprintTitle{%
 \let\@oddhead\@empty
 \let\@evenhead\@empty
 \def\@oddfoot{}%
 \let\@evenfoot\@oddfoot}
\makeatother
\usepackage{graphicx}
\usepackage{amsbsy}
\usepackage[usenames, dvipsnames]{color}

\usepackage[utf8x]{inputenc}
\usepackage{epstopdf}
\usepackage{amsmath,amssymb}

\usepackage{lineno,hyperref}
\begin{document}
\title{Experimental detection of nonclassicality of single-mode fields via intensity moments}

\author[slo]{Ievgen I. Arkhipov}
\address[slo]{RCPTM, Joint Laboratory of Optics of Palack\'y University and
Institute of Physics of CAS, Faculty of Science, Palack\'y University, 17. listopadu
12, 771 46 Olomouc, Czech Republic}
\author[fzu]{Jan Pe\v{r}ina Jr.\corref{cor}}
\ead{jan.perina.jr@upol.cz}
\address[fzu]{Institute of Physics of the Czech Academy of Sciences, Joint
Laboratory of Optics of Palack\'{y} University and Institute of
Physics of CAS, 17. listopadu 50a, 772 07 Olomouc, Czech Republic}
\cortext[cor]{Corresponding author}
\author[slo]{Ond\v{r}ej Haderka}
\author[fzu]{V\'{a}clav Mich\'{a}lek}


\begin{abstract}
Nonclassicality criteria based on intensity moments and derived
from the usual matrix approach are compared to those provided by
the majorization theory. The majorization theory is shown to give
a greater number of more suitable nonclassicality criteria.
Fifteen experimentally useful criteria of the majorization theory
containing the intensity moments up to the fifth order are
identified. Their performance is experimentally demonstrated on
the set of eleven potentially nonclassical states generated from a
twin beam by postselection based on detecting a given number of
photocounts in one arm by using an iCCD camera.
\end{abstract}
\maketitle


\section{Introduction}

The world of quantum states of optical fields was discovered soon
after the construction of the first laser that opened the area of
nonlinear optics \cite{Bloembergen1965} for extensive
investigations. The experimental effort has been supported by the
fast development of theory \cite{Glauber2007} that provided soon
the crucial coherent states, the Glauber-Sudarshan representation
of the statistical operator and quantum generalization of the
coherence theory \cite{Glauber1963,Sudarshan1963}. The
Glauber-Sudarshan representation also brought a clear and strict
definition of nonclassical optical states \cite{Glauber1963}: A
quantum state of an optical field is nonclassical if and only if
the Glauber-Sudarshan representation of its statistical operator
fails to be a probability density (it attains negative values or
even does not exist as a regular function). Applying this
definition, a huge amount of quantum states can be identified.
From the point of view of the fundamental physical importance,
three distinct kinds of such states have attracted the greatest
attention of experimentalists from the very beginning
\cite{Mandel1995}: squeezed states with reduced phase
fluctuations, sub-Poissonian states with reduced intensity (or
photon-number) fluctuations and anti-bunched light with unusual
temporal correlations.

The verification of nonclassicality of an optical state can be
done directly from its definition given in the above paragraph,
provided that its statistical operator is reconstructed from the
measured data. However, this requires in general the application
of homodyne tomography \cite{Lvovsky2009} or homodyne detection
\cite{Shchukin2006}, which is experimentally demanding.
Qualitative simplification is reached only for optical fields with
the uniform distribution of phases for which the measurement of
photocount statistics is sufficient to reconstruct the
quasi-distribution of integrated intensities
\cite{Perina1991,Haderka2005a,PerinaJr2008,PerinaJr2012,PerinaJr2013a}
that fully characterizes the field. We note that even the full
state reconstruction can be reached using only the photocount
statistics in some specific cases, e.g., for a two-mode Gaussian
field without coherent contributions \cite{arkhipov2016a}. Because
of the complexity of state reconstruction, alternative approaches
for revealing the nonclassicality of a state have been looked for,
even considering transformations of the nonclassicality into its
different forms \cite{Asboth05,Arkhipov2016,Arkhipov2016b}.
 A large number of various inequalities comprising both moments of
amplitudes and intensities of different orders have been derived
\cite{Lee1990,Lee1990b,Verma2010,Sperling2012a,Giri2014} and
experimentally tested
\cite{Short1983,Avenhaus2010,Allevi2012a,Sperling2015}. A unifying
matrix approach for their derivation has been formulated relying
on nonnegativity of classical quadratic forms defined above
amplitude and intensity powers of different orders
\cite{Agarwal1992,Shchukin2005,Vogel2008}. Different inequalities
have been compared in \cite{Miranowicz2010}. Even more general
forms of such inequalities have been reached applying the Bochner
theorem \cite{Richter2002,Ryl2015}. There also exists a completely
different approach for the derivation of such inequalities based
on the mathematical theory of majorization
\cite{Marshall2010,Lee1990a}.

Inequalities containing only moments of intensities are frequently
used to reveal the nonclassicality of experimentally investigated
states, contrary to those written for amplitude moments. This is
natural, as the measurement of amplitude moments requires the
homodyne scheme \cite{Mandel1995} whose complexity of
implementation is comparable to the homodyne tomography. On the
other hand, intensity moments can be obtained with the usual
'quadratic' optical detectors or, for low intensities, with their
modern variants resolving individual photon numbers
\cite{Allevi2010}. In this contribution, we compare the
nonclassicality inequalities derived from the matrix approach with
those provided by the majorization theory using a set of
sub-Poissonian states with increasing mean photon numbers
\cite{PerinaJr2013b}. These states are generated from a twin beam
\cite{Arkhipov2015} by postselection
\cite{Laurat2003,Lamperti2014} that is based on the detection of a
given photocount number in one arm of the twin beam by an
intensified charge-coupled-device (iCCD) camera \cite{Hamar2010}.
The iCCD camera is also used to experimentally analyze the
sub-Poissonian states with mean photon numbers ranging from 7 to
14.

The paper is organized as follows. In Sec.~2, systematic approach
for the derivation of nonclassicality inequalties is given using
both the matrix approach and the majorization theory. Inequalities
derived in Sec.~2 are tested on the experimental data in Sec.~3.
Sec.~4 brings conclusions.

\section{Derivation of nonclassicality inequalities}

For the moments of classical integrated intensity $ I $
\cite{Perina1991}, a general nonnegative quadratic form for the
classical field is constructed via the function $ g(I) $ that is
an arbitrary linear superposition of the terms $ I^j $ for $
j=0,1,\ldots $:
\begin{equation} 
 g(I) = \sum_{j=0}^{N} g_j I^j ,
\label{1}
\end{equation}
and $ N $ is an arbitrary integer number giving the number of
terms in the sum. The condition $ \int_{0}^{\infty} dI\, P(I)
|g(I)|^2 \ge 0 $ for a classical state with { non-negative}
probability function $ P $, when transformed into the operator
form written for the powers of photon-number operator $ \hat{n} $
($ I^j \propto $ $:\hat{n}^j\!: $), suggests the following
nonclassicality condition
\cite{Agarwal1992,Shchukin2005,Vogel2008,Miranowicz2010}:
\begin{equation} 
 \sum_{j,j'=0}^N g_j g_{j'} \langle:\!\hat{n}^{j+j'}\!:\rangle <
 0 ;
\label{2}
\end{equation}
symbol $:\::$ denotes normal ordering of field operators.
Inequality (\ref{2}) can be equivalently expressed as the
condition for negativity of a matrix $ { M} $ of dimension $
(N+1) \times (N+1) $ with the elements $ M_{jj'} =
\langle:\!\hat{n}^{j+j'}\!:\rangle $. The Hurwitz criterion then
guarantees negativity of the matrix $ { M} $ whenever any of
its principal minors is negative.

The simplest $ 2\times 2 $ minors of the matrix $ { M} $
written as
\begin{equation}       
 {\rm det } \left[ \begin{array}{cc} \langle:\!\hat n^{2k}\!:\rangle & \langle:\!\hat n^{k+l}\!:\rangle
 \cr \langle:\!\hat n^{k+l}\!:\rangle & \langle:\!\hat n^{2l}\!:\rangle \end{array}
 \right]
\label{3}
\end{equation}
provide the nonclassicality inequalities containing the products
of two moments of in general different orders:
\begin{equation}    
 \langle:\!\hat n^{2k}\!:\rangle \langle:\!\hat n^{2l}\!:\rangle <
  \langle:\!\hat n^{k+l}\!:\rangle^2, \hspace{5mm} 0 \le k \le l.
\label{4}
\end{equation}

The $ 3\times 3 $ minors of matrix $ { M} $ parameterized by
integer numbers $ k $, $ l $ and $ m $ already give more complex
nonclassicality inequalities involving in general 6 terms in the
sum, each formed by three moments in the product:
\begin{eqnarray}    
 {\rm det}\left[ \begin{array}{ccc}  \langle:\!\hat n^{2k}\!:\rangle & \langle:\!\hat n^{k+l}\!:\rangle
  & \langle:\!\hat n^{k+m}\!:\rangle \cr
  \langle:\!\hat n^{k+l}\!:\rangle & \langle:\!\hat n^{2l}\!:\rangle & \langle:\!\hat n^{l+m}\!:\rangle \cr
  \langle:\!\hat n^{k+m}\!:\rangle & \langle:\!\hat n^{l+m}\!:\rangle & \langle:\!\hat n^{2m}\!:\rangle \cr
  \end{array}\right]
  <0, \hspace{5mm} 0 \le k \le l \le m .
\label{5}
\end{eqnarray}
The form of nonclassicality inequalities originating in $ k\times
k $ minors for $ k>3 $ is similar to that derived for the $
3\times 3 $ minors.

On the other hand, the majorization theory \cite{Marshall2010}
gives us the nonclassicality inequalities involving two moments in
the product and having the following form~\cite{Lee1990a}:
\begin{eqnarray}       
 \tilde{R}_{u,v}^{u+m,v-m} \equiv \langle:\!\hat n^{u+m}\!:\rangle
  \langle:\!\hat n^{v-m}\!:\rangle - \langle:\!\hat n^{u}\!:\rangle
  \langle:\!\hat n^{v}\!:\rangle < 0, \hspace{5mm} u \ge v \ge 0, \; v \ge m \ge 0. \label{6}
\end{eqnarray}
The inequalities written in Eq.~(\ref{4}) are a subset of those
given in Eq.~({\ref{6}) with the mapping $ u=v=k+l $ and $ m = l-k
$.

The following nonclassicality inequalities of the majorization
theory represent the counterpart of inequalities in Eq.~(\ref{6})
derived from the $ 3\times 3 $ minors:
\begin{eqnarray}\label{7}  
 \tilde{R}_{u,v,w}^{u+k+l,v-k+m,w-l-m} \hspace{-3mm} & \equiv &\hspace{-3mm}
  \langle:\!\hat n^{u+k+l}\!:\rangle \langle:\!\hat n^{v-k+m}\!:\rangle
  \langle:\!\hat n^{w-l-m}\!:\rangle - \langle:\!\hat n^{u}\!:\rangle
  \langle:\!\hat n^{v}\!:\rangle \langle:\!\hat n^{w}\!:\rangle <0, \nonumber  \\
   & & \hspace{5mm} u \ge v \ge w \ge 0; \hspace{5mm} \; k,l,m \ge 0.
\end{eqnarray}
Inequalities in Eq.~(\ref{7}) formed by two additive terms differ
from those of Eq.~(\ref{5}) that contain six additive terms. There
does not seem to exist any simple relation between inequalities
written in Eqs.~(\ref{7}) and (\ref{5}).

In general, the majorization theory provides a larger number of
nonclassicality inequalities compared to the matrix approach.
Moreover these inequalities attain a simpler form [compare
Eqs.~(\ref{5}) and (\ref{7})]. To get a more detailed comparison
of the two methods, we write down explicitly the inequalities
involving the moments with the overall power up to five, that are
useful for the experimental analysis below. The explicit formulas
of Eq.~(\ref{6}) written in their normalized (dimensionless) form
are expressed as follows:
\begin{eqnarray}       
 R_{1,1}^{2,0} &\equiv & \frac{\langle:\!\hat n^{2}\!:\rangle}{\langle:\!\hat n\!:\rangle^{2}}
   - 1 < 0, \nonumber  \\
 R_{2,1}^{3,0} &\equiv & \frac{\langle:\!\hat n^{3}\!:\rangle}{\langle:\!\hat n\!:\rangle^{3}}
   - \frac{\langle:\!\hat n^2\!:\rangle}{\langle:\!\hat n\!:\rangle^{2}} < 0, \nonumber  \\
 R_{2,2}^{3,1} &\equiv & \frac{\langle:\!\hat n^{3}\!:\rangle}{\langle:\!\hat n\!:\rangle^{3}}
   - \frac{\langle:\!\hat n^2\!:\rangle^2}{\langle:\!\hat n\!:\rangle^{4}} < 0, \nonumber  \\
 R_{2,2}^{4,0} &\equiv & \frac{\langle:\!\hat n^{4}\!:\rangle}{\langle:\!\hat n\!:\rangle^{4}}
  - \frac{\langle:\!\hat n^2\!:\rangle^2}{\langle:\!\hat n\!:\rangle^{4}} < 0, \nonumber \\
 R_{3,1}^{4,0} &\equiv & \frac{\langle:\!\hat n^{4}\!:\rangle}{\langle:\!\hat n\!:\rangle^{4}}
   - \frac{\langle:\!\hat n^3\!:\rangle}{\langle:\!\hat n\!:\rangle^{3}} < 0, \nonumber  \\
 R_{4,1}^{5,0} &\equiv & \frac{\langle:\!\hat n^{5}\!:\rangle}{\langle:\!\hat n\!:\rangle^{5}}
  - \frac{\langle:\!\hat n^4\!:\rangle}{\langle:\!\hat n\!:\rangle^{4}} < 0, \nonumber  \\
 R_{3,2}^{4,1} &\equiv & \frac{\langle:\!\hat n^{4}\!:\rangle}{\langle:\!\hat n\!:\rangle^{4}}
   - \frac{\langle:\!\hat n^2\!:\rangle \langle:\!\hat n^3\!:\rangle}{\langle:\!\hat n\!:\rangle^{5}} < 0, \nonumber  \\
 R_{3,2}^{5,0} &\equiv & \frac{\langle:\!\hat n^{5}\!:\rangle}{\langle:\!\hat n\!:\rangle^{5}}
   - \frac{\langle:\!\hat n^2\!:\rangle \langle:\!\hat n^3\!:\rangle}{\langle:\!\hat n\!:\rangle^{5}} < 0.
\label{8}
\end{eqnarray}
Only the first and the fourth inequalities in Eq.~(\ref{8}) stem
from the matrix approach providing Eq.~(\ref{4}). Similarly, the
general formula in Eq.~(\ref{7}) leaves us with the following four
normalized inequalities, which cannot be obtained from the matrix
approach:
\begin{eqnarray}       
 R_{1,1,1}^{3,0,0} &\equiv & \frac{\langle:\!\hat n^{3}\!:\rangle}{\langle:\!\hat n\!:\rangle^{3}}
   - 1 < 0, \nonumber  \\
 R_{2,1,1}^{4,0,0} &\equiv & \frac{\langle:\!\hat n^{4}\!:\rangle}{\langle:\!\hat n\!:\rangle^{4}}
   - \frac{\langle:\!\hat n^2\!:\rangle}{\langle:\!\hat n\!:\rangle^{2}} < 0, \nonumber  \\
 R_{2,2,1}^{5,0,0} &\equiv & \frac{\langle:\!\hat n^{5}\!:\rangle}{\langle:\!\hat n\!:\rangle^{5}}
   - \frac{\langle:\!\hat n^2\!:\rangle^2}{\langle:\!\hat n\!:\rangle^{4}} < 0, \nonumber  \\
 R_{3,1,1}^{5,0,0} &\equiv & \frac{\langle:\!\hat n^{5}\!:\rangle}{\langle:\!\hat n\!:\rangle^{5}}
   - \frac{\langle:\!\hat n^3\!:\rangle}{\langle:\!\hat n\!:\rangle^{3}} < 0.
\label{9}
\end{eqnarray}

The majorization theory gives us also inequalities containing four
(five) moments in the product, which encompass the following two
(one) inequalities useful in our experimental analysis:
\begin{eqnarray}   
 R_{1,1,1,1}^{4,0,0,0} &\equiv & \frac{\langle:\!\hat n^4\!:\rangle}{\langle:\!\hat n\!:\rangle^{4}}
   - 1 < 0, \nonumber \\
 R_{2,1,1,1}^{5,0,0,0} &\equiv & \frac{\langle:\!\hat n^5\!:\rangle}{\langle:\!\hat n\!:\rangle^{5}}
   - \frac{\langle:\!\hat n^2\!:\rangle}{\langle:\!\hat n\!:\rangle^{2}}  < 0, \nonumber \\
  R_{1,1,1,1,1}^{5,0,0,0,0} &\equiv & \frac{\langle:\!\hat n^5\!:\rangle}{\langle:\!\hat n\!:\rangle^{5}}
   - 1 < 0.
\label{10}
\end{eqnarray}

The above inequalities can be applied both to photon-number
distributions as well as to photocount distributions that are
directly measured. Whereas the normally-ordered moments of photon
number $ \hat n $ are suitable for characterizing intensity
distributions, the usual moments are immediately derived from the
experimental photocount distributions. They are mutually related
by the following formula \cite{Perina1991}:
\begin{equation}        
 \langle:\!\hat{n}^k\!:\rangle= \left\langle \frac{\hat{n}!}{(\hat{n}-k)!}\right\rangle .
\label{11}
\end{equation}
\begin{figure}  
 \centerline{\resizebox{0.8\hsize}{!}{\includegraphics{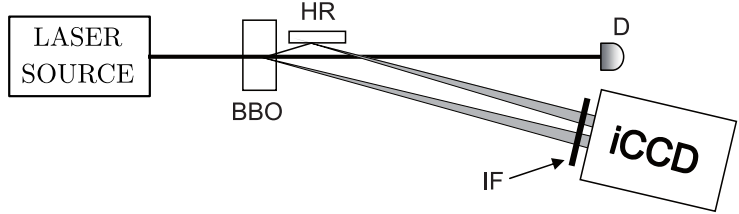}}}
 \caption{Scheme of the experiment generating sub-Poissonian
 states. A twin beam is generated in noncollinear geometry in a
 5-mm-long type-I barium-borate crystal (BaB$ {}_2 $O$ {}_4 $, BBO) pumped by the third
 harmonics (280~nm) of a femtosecond cavity dumped Ti:sapphire laser (pulse
 duration 150~fs, central wavelength 840~nm). The signal field as
 well as the idler field (after reflection on a highly-reflecting mirror HR) are
 detected by $N_{\rm s}=6528$ and $N_{\rm i}=6784$ pixels of the photocathode
 of iCCD camera Andor DH3345-18U-63 with dark-count rate $ d = 0.04 $ ($ D_a=d/N_a$, $ a={\rm s,i} $).
 The nearly-frequency-degenerate signal and
 idler photons at the wavelength of 560~nm are filtered by
 a 14-nm-wide bandpass interference filter IF. Intensity of the pump beam that is actively
 stabilized via a motorized half-wave plate followed by a polarizer is monitored by
 detector D.}
\end{figure}

\section{Experimental testing of nonclassicality inequalities}

The above nonclassicality inequalities have been applied to a set
of states with different 'degree' of sub-Poissonian photon-number
statistics that were generated from a twin beam using
postselection based on the detection of a given number $ c_{\rm s}
$ of photocounts in the signal field \cite{PerinaJr2013b}. For an
ideal photon-number-resolving detector, detection of a given
number $ c_{\rm s} $ of signal photocounts leaves the idler field
in the Fock state with $ c_{\rm s} $ photons. For a real
photon-number-resolving detector, an idler field with reduced
photon-number fluctuations and potentially sub-Poissonian
photon-number statistics is obtained. As the mean number of idler
photons in a postselected field increases with the increasing
signal photocount number $ c_{\rm s} $, the set of generated
states is appealing for testing the power of the nonclassicality
inequalities.

In the reported experiment, the twin beam was generated in a
nonlinear crystal and both its signal and idler fields were
detected by an iCCD camera \cite{PerinaJr2012}. Whereas the signal
photocounts were used for the postselection process, the
histograms of idler photocounts provided the information about the
postselected potentially sub-Poissonian idler fields. Experimental
details are written in the caption to Fig.~1. The experiment was
repeated $1.2\times 10^6$ times. The obtained 2D histogram of the
signal and idler photocounts was used both to determine the
photocount moments occurring in the nonclassicality inequalities
and to reconstruct the photon-number distributions of the
postselected idler fields by the method of maximum likelihood
\cite{Dempster1977}. In the reconstruction method, the idler-field
conditional photon-number distribution $ p_{\rm c,i}(n_{\rm
i};c_{\rm s}) $ left after detecting $ c_{\rm s} $ signal
photocounts is reached as a steady state found in the following
iteration procedure (with iteration index $ n $)
\cite{PerinaJr2012}
\begin{equation} 
 p_{\rm c,i}^{(n+1)}(n_{\rm i};c_{\rm s}) = p_{\rm c,i}^{(n)}(n_{\rm i};c_{\rm s})
  \sum_{c_{\rm i}} \frac{ f_{\rm i}(c_{\rm i};c_{\rm s})
  T_{\rm i}(c_{\rm i},n_{\rm i}) }{ \sum_{n'_i} T_{\rm i}(c_{\rm i},n'_{\rm i})
  p_{\rm c,i}^{(n)}(n'_{\rm i};c_{\rm s}) }
\label{12}
\end{equation}
that uses the normalized idler-field 1D photocount histogram $f_{\rm i}(c_{\rm i};c_{\rm s}) \equiv \frac{f(c_{\rm s},c_{\rm i})}
{\sum\limits_{c_{\rm i}}f(c_{\rm s},c_{\rm i})}$ built from the detections
with $ c_{\rm s} $ detected signal photocounts and contained in
the joint signal-idler photocount histogram $ f(c_{\rm s},c_{\rm
i}) $. In Eq.~(\ref{12}), the functions $ T(c_{\rm i},n_{\rm i}) $
give the probabilities of having $ c_{\rm i} $ photocounts when
detecting a field with $ n_{\rm i} $ photons. The folowing formula
was derived for an iCCD camera with $ N_a $ active pixels,
detection efficiency $ \eta_a $ and dark-count rate per pixel $
D_a $ \cite{PerinaJr2012}:
\begin{eqnarray}     
  T_a(c_a,n_a) &=& \left(\begin{array}{c} N_a \\ c_a \end{array}\right) (1-D_a)^{N_a}
  (1-\eta_a)^{n_a} (-1)^{c_a} \sum_{l=0}^{c_a}
  \left(\begin{array}{c} c_a \\ l \end{array}\right) \frac{(-1)^l}{(1-D_a)^l} \nonumber \\
   & &  \mbox{} \times  \left( 1 + \frac{l}{N_a} \frac{\eta_a}{1-\eta_a}
   \right)^{n_a}; \hspace{10mm} a={\rm s,i}.
\label{13}
\end{eqnarray}

The 2D histogram $ f(c_{\rm s},c_{\rm i}) $ with $\langle c_{\rm
s}\rangle=2.20$ and $\langle c_{\rm i}\rangle=2.18$ signal and
idler mean photocounts, respectively, also allowed to reconstruct
the whole original twin beam in the form of multimode Gaussian
fields composed of the independent multimode paired, signal and
idler parts characterized by mean photon(-pair) numbers $ B_a $
per mode and numbers $ M_a $ of independent modes, $ a={\rm p,s,i}
$ \cite{PerinaJr2013a}. The photon-nunber distribution $ p_{\rm
si}(n_{\rm s},n_{\rm i}) $ of the whole twin beam was expressed in
the form of a two-fold convolution of three Mandel-Rice
photon-number distributions \cite{Perina1991} in this case
\cite{PerinaJr2012a,PerinaJr2013a,Perina2005}:
\begin{eqnarray}  
 p_{\rm si}(n_{\rm s},n_{\rm i}) = \sum_{n=0}^{{\rm min}[n_{\rm s},n_{\rm i}]} p(n_{\rm s}-n;M_{\rm s},B_{\rm s})
  p(n_{\rm i}-n;M_{\rm i},B_{\rm i}) p(n;M_{\rm p},B_{\rm p});
\label{14}
\end{eqnarray}
$ p(n;M,B) = \Gamma(n+M) / [n!\, \Gamma(M)] b^n/(1+B)^{n+M} $ and
symbol $ \Gamma $ stands for the $ \Gamma $-function. This
reconstruction revealed the following values of mean photon(-pair)
numbers $ B_a $ and numbers $ M_a $ of modes: $M_{\rm p}=270$,
$B_{\rm p}=0.032$, $M_{\rm s}=0.01$, $B_{\rm s}=7.6$, $M_{\rm i} =
0.026$, and $B_{\rm i}=5.3$. The method also provided the signal
($ \eta_{\rm s}=0.23 $) and idler ($ \eta_{\rm i}=0.22 $)
detection efficiencies and the theoretical prediction for the
conditional idler-field photon-number distributions $ p_{\rm
c,i}^{\rm t}(n_{\rm i};c_{\rm s}) $ arising in the postselection
process with $ c_{\rm s} $ detected signal photounts (for details,
see \cite{PerinaJr2013b}):
\begin{equation}   
 p_{\rm c,i}^{\rm t}(n_{\rm i};c_{\rm s}) = \frac{
  \sum_{n_{\rm s}} T_{\rm s}(c_{\rm s},n_{\rm s}) p_{\rm si}(n_{\rm s},n_{\rm i}) }{
  f_{\rm s}^{\rm t}(c_{\rm s}) }
\label{15}
\end{equation}
where $ f_{\rm s}^{\rm t}(c_{\rm s}) \equiv \sum_{n_{\rm s},
n_{\rm i}} T_{\rm s}(c_{\rm s},n_{\rm s}) p_{\rm si}(n_{\rm
s},n_{\rm i}) $ is the theoretical prediction for the signal-field
photocount distribution.

In the experiment, eleven conditional idler fields generated after
detection of a given number $ c_{\rm s} $ of signal photocounts in
the range $ <0,10> $ were analyzed. Their mean photocount numbers
$ \langle c_{\rm i} \rangle $ and photon numbers $ \langle n_{\rm
c,i} \rangle $ plotted in Fig.~2(a) show that the conditional
idler fields contained from 7 to 14 photons on average.
\begin{figure}  
 \centerline{\includegraphics[width=0.45\textwidth]{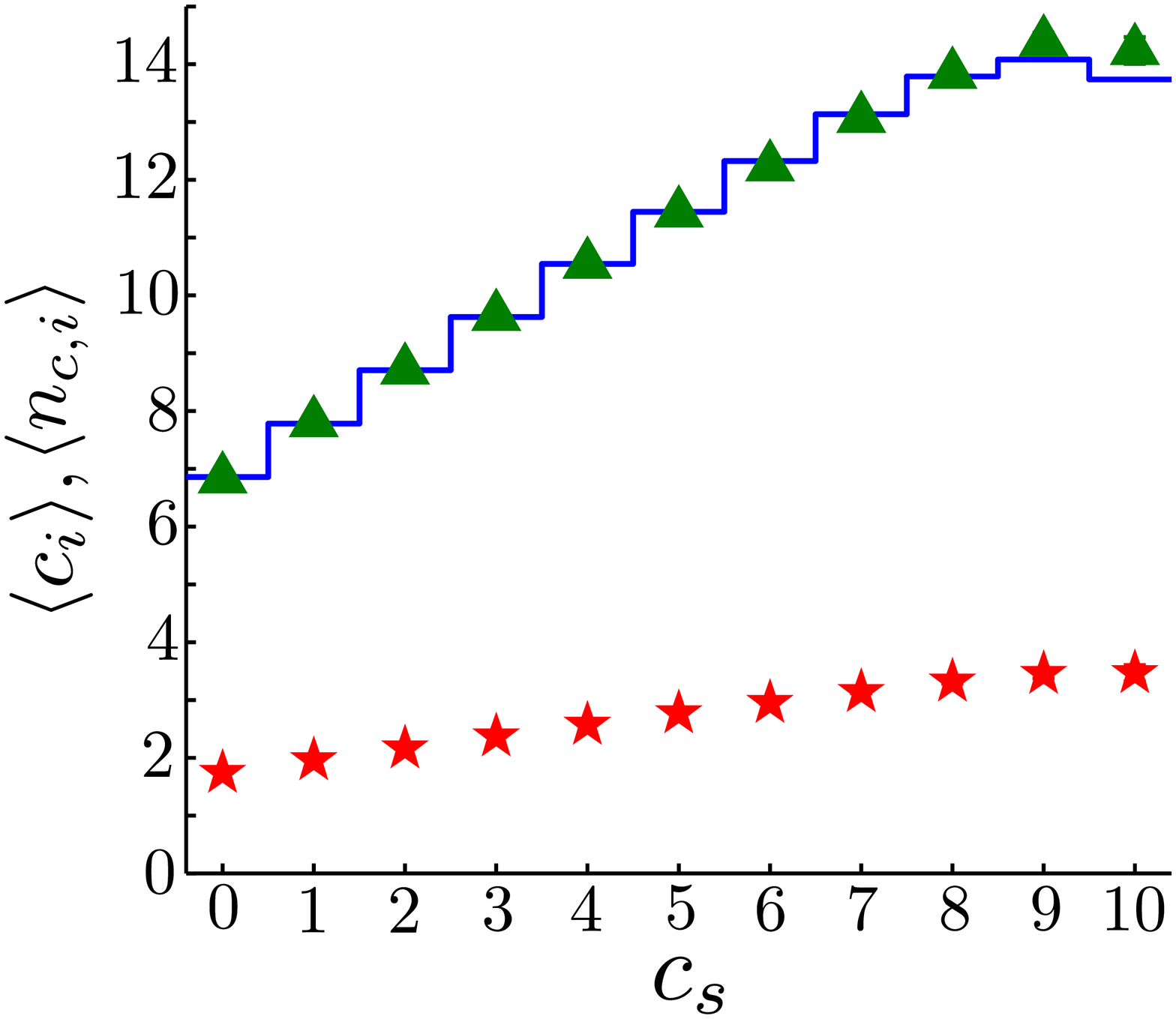}
 \hspace{5mm}
 \includegraphics[width=0.45\textwidth]{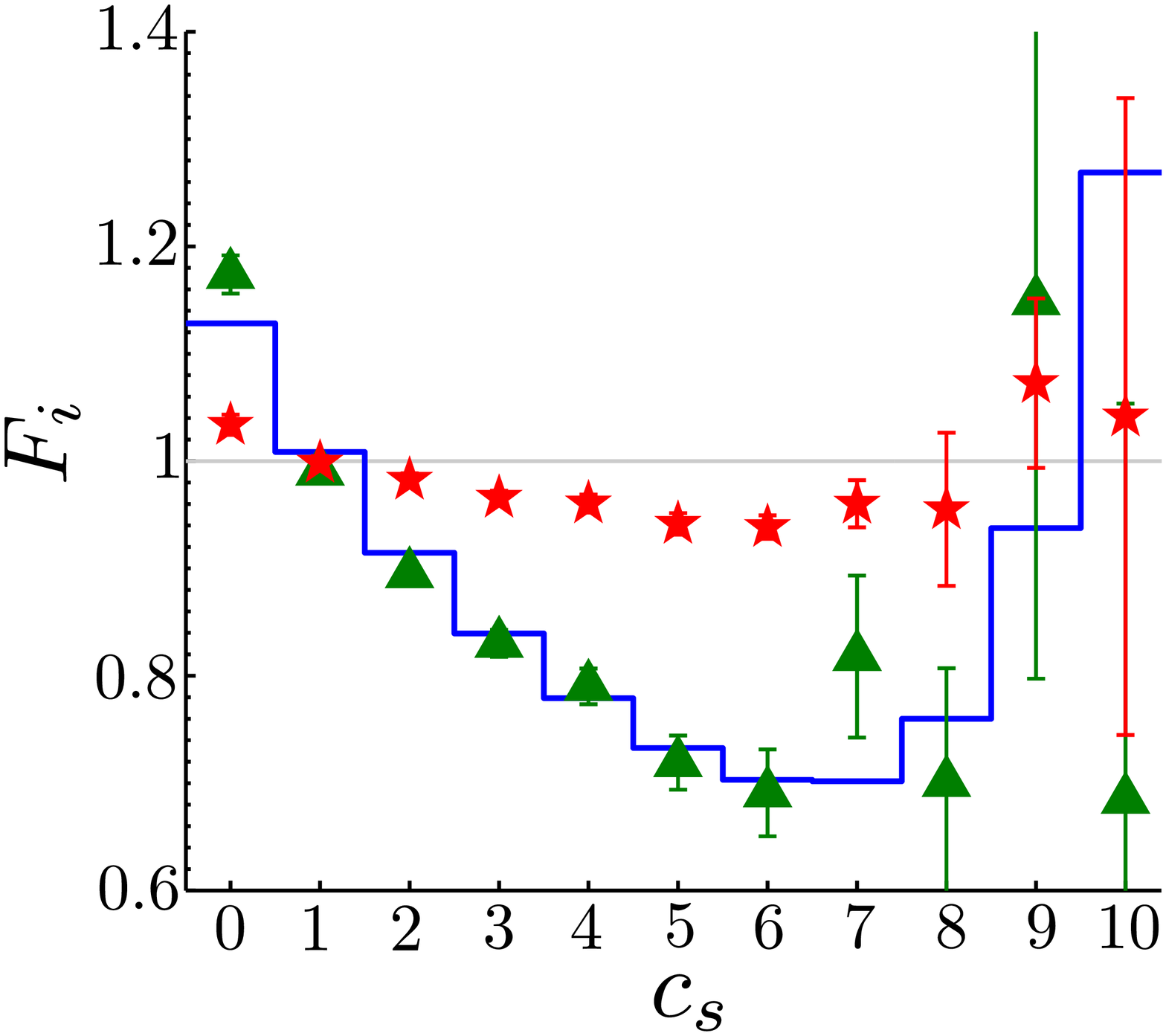}}
 \centerline{(a) \hspace{.5\textwidth} (b)}
 \caption{(a) Mean idler photocount number $ \langle c_{\rm i} \rangle $ and
  photon number $ \langle n_{\rm c, i} \rangle $ and (b) Fano factor $ F_{\rm i} $
  [$ F_{\rm i} \equiv (\langle \hat n_{\rm i}^2\rangle - \langle \hat n_{\rm i}\rangle^2)/
  \langle \hat n_{\rm i}\rangle $] as they depend on
  the signal photocount number $ c_{\rm s} $. The values appropriate for
  the distributions of experimental photocounts are plotted with red asterisks whereas those
  characterizing the reconstructed photon-number distributions arising from the
  maximum-likelihood method (from the best fit of the twin beam)
  are plotted with green triangles (blue solid curves).
  Error bars in (a) are smaller than the used symbols.}
\end{figure}
The corresponding Fano factors $ F_{\rm i} $ determined from the
first- and second-order moments and drawn in Fig.~2(b) identify,
within the experimental errors, the conditional fields with $
c_{\rm s} \in <2,7> $ as sub-Poissonian. They also suggest that
the nonclassicality of the conditional idler fields increases as $
c_{\rm s} $ increases from 2 to 6, but then the nonclassicality
decreases and it is lost for $ c_{\rm s} = 9 $. This behavior
originates in the noise present both in the experimental twin beam
and the iCCD camera (that makes the postselection, smaller $
c_{\rm s} $) as well as the relatively low detection efficiency of
the iCCD camera (greater $ c_{\rm s} $) \cite{PerinaJr2013b}.

Fano factors $ F_{\rm i } $ for $ c_{\rm s} >4 $ plotted in
Fig.~2(b) are determined with larger errors that increase with the
increasing signal photocount number $ c_{\rm s} $. This originates
in relatively small numbers of measurements appropriate for the
mentioned numbers $ c_{\rm s} $. Mean numbers of these
measurements are described by the signal-field photocount
distribution $ f_{\rm s} $ plotted in Fig.~3.
\begin{figure}  
 \centerline{\includegraphics[width=0.5\textwidth]{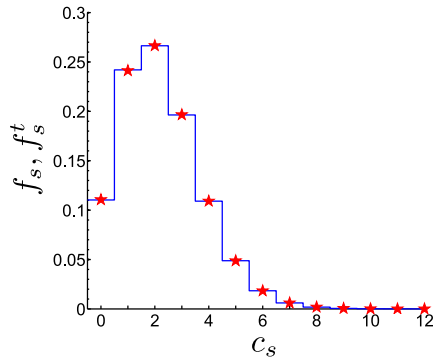}}
 \caption{Marginal signal-field photocount distribution
  $ f_{\rm s}(c_{\rm s}) = \sum_{c_{\rm i}} f(c_{\rm s},c_{\rm i}) $ (red asterisks) and
  its theoretical prediction $ f^{\rm t}_{\rm s} $ defined below Eq.~(\ref{15}) (blue solid
  curve). Error bars of $ f_{\rm s} $ are smaller than the used symbols.}
\end{figure}
According to this distribution, the probability of detecting the
signal photocount numbers $ c_{\rm s} $ greater than 7 is less
than 1~\%. Despite the large number $ N=1.2\times 10^6 $ of
experimental repetitions, the determined quantities suffer from
relatively large experimental errors in these cases. The
experimental errors (for photocounts) are quantified by the mean
squared fluctuation $ \sigma_x $ ($ \sigma_x = \sqrt{ \langle
x^2\rangle - \langle x\rangle^2 } $) multiplied by factor $
1/\sqrt{N_{\rm r}} $ that depends on the number $ N_{\rm r} $ of
actual experimental realizations. This approach was also applied
to the determination of error bars of the quantities
characterizing the photon-number distributions reached by the
maximum-likelihood reconstruction. In this method that gives the
most-probable photon-number distribution the experimental errors
are naturally smoothed [also due to the form of $ T_a $ given in
Eq.~(\ref{13}) that includes $ D_a $]. We note that the extended
approach based on the Fischer information matrix
\cite{Frieden2004} allows to quantify their contribution to the
uncertainty characterizing the reconstructed photon-number
distribution. The reconstruction based on the best fit of the 2D
experimental histogram $ f(c_{\rm s},c_{\rm i}) $ exploits the
whole ensemble of the measured data with $ N=1.2\times 10^6 $
entries and so the corresponding relative errors are negligible.

The nonclassicality identifiers $ R $ belonging to the
experimental photocounts and plotted in Fig.~4 identify the
conditional idler fields with $ c_{\rm s} \in <2,7> $ as
nonclassical, in agreement with the predictions made by the Fano
factors.
\begin{figure}  
 \includegraphics[width=0.98\textwidth]{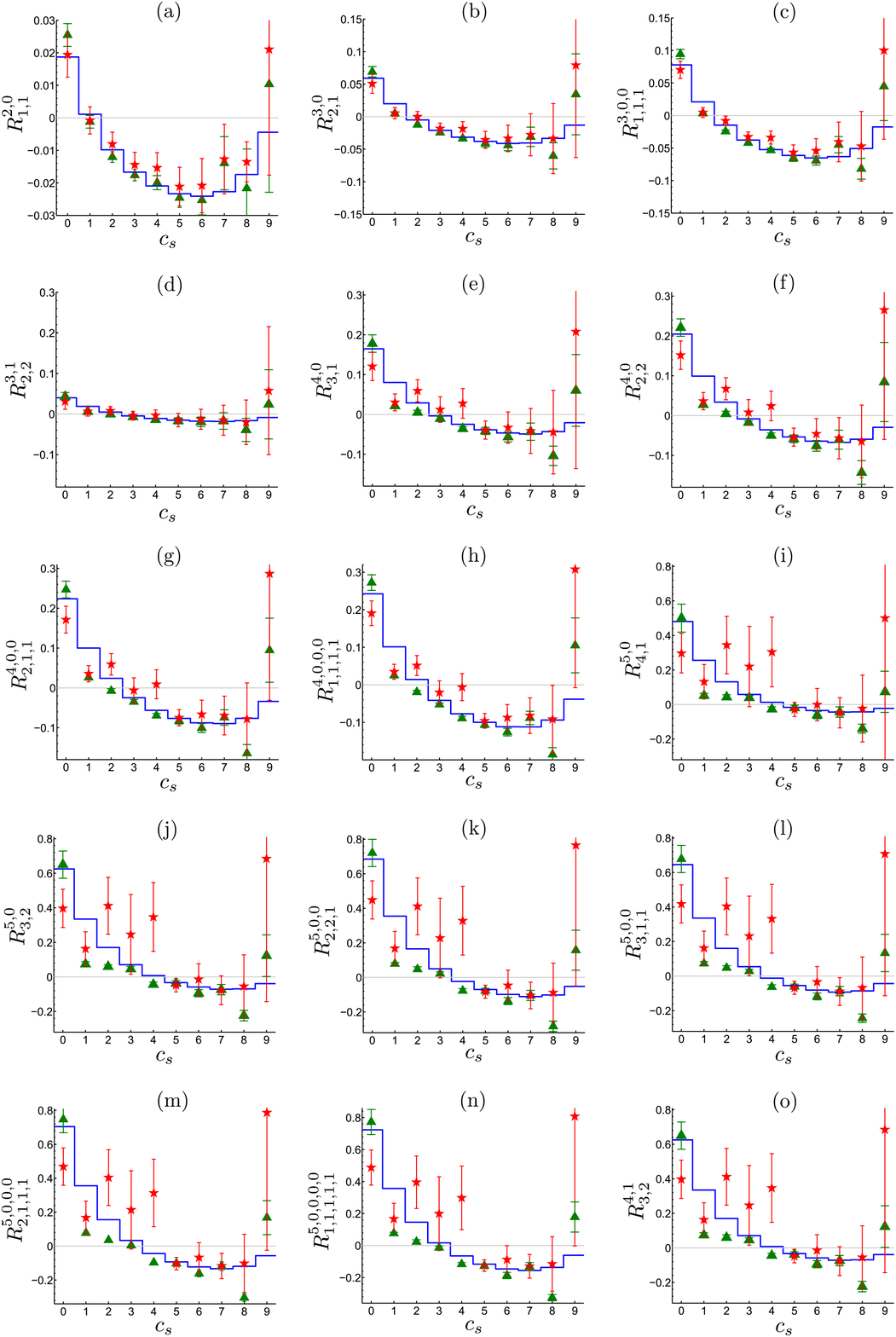}
 \caption{Nonclassicality identifiers $  R  $
  given in Eqs.~(\ref{8})---(\ref{10}) and determined for
  distributions of experimental photocounts (red asterisks with error bars), photon-number distributions
  reached by the maximum-likelihood reconstruction method (green triangles with error bars) and photon-number distributions
  derived from the best fit of the twin beam (blue solid curves) as they depend on
  the signal photocount number $ c_{\rm s} $. Some error bars are smaller than the plotted symbols.}
\end{figure}
According to the graphs in Fig.~4, nonclassical conditional idler
fields can be divided into two groups. The conditional idler
fields with $ c_{\rm s} \in $ \mbox{$ <5,7> $} have all fifteen
nonclassicality identifiers $ R $ negative, determined both for
photocounts and photon numbers. Such fields can thus be considered
as firmly nonclassical. This accords with the lowest attained
values of the Fano factor $ F_{\rm i}(c_{\rm s}) $ shown in
Fig.~2(b). On the other hand, the conditional idler fields with $
c_{\rm s} \in <2,4> $ have negative only the nonclassicality
identifiers $ R $ of the 'order', given as the sum of their upper
(or equivalently lower) indices, lower than four. The
nonclassicality identifiers $ R $ of 'order' four and five are
positive for the experimental photocounts. The maximum-likelihood
reconstruction, that relies on the whole 1D experimental
histograms, additionally provides negative nonclassicality
identifiers $ R $ of 'order' four for $ c_{\rm s} \in $
\mbox{$<3,4> $} and five for $ c_{\rm s} = 4 $. This corresponds
to the decreasing values of Fano factor $ F_{\rm i}(c_{\rm s}) $
drawn in Fig.~2(b). Detailed inspection of the graphs in Fig.~4
reveals that the behavior of nonclassicality identifiers $ R $
reached by the maximum-likelihood reconstruction qualitatively
agrees (up to $ c_{\rm s} = 7 $) with the behavior predicted by
the reconstruction based on the best fit of the 2D experimental
histogram and quantified by blue solid curves in the graphs of
Figs.~2 and 4.

Compared to the Fano factor $ F_{\rm i} $, the nonclassicality
identifiers $ R $ of 'order' three or higher are endowed with
weaker capability to reveal the nonclassicality of the analyzed
states obtained by the post-selection method. The greater the
'order' of nonclassicality identifier $ R $ the weaker the
capability. On the other hand, if the nonclassicality is observed
in the nonclassicality identifiers $ R $ of higher 'order' it can
be considered in certain sense as firm. This is the case of the
conditional idler fields obtained after postselecting by the
detection of 5, 6 and 7 signal photocounts. These fields,
containing on average about 12-14 idler photons, exhibit their
nonclassicality in all observed nonclassicality identifiers.

\section{Conclusions}

We have shown that the majorization theory provides a greater
number and more suitable nonclassicality identifiers based on
intensity moments compared to the commonly used matrix method.
Considering the products of moments up to the fifth order, we
identified fifteen independent identifiers and tested them on the
experimental states with different 'degree' of sub-Poissonian
photon-number statistics. Identifiers based on lower intensity
moments were identified as more powerful compared to those
containing greater intensity moments. The latter ones have been
found useful for identifying states being firmly nonclassical.

\section*{Funding}

GA \v{C}R (project P205/12/0382); M\v{S}MT \v{C}R (project
LO1305); IGA UP Olomouc (project IGA\_PrF\_2016\_002) - I.
Arkhipov.

\section*{Acknowledgments}
The authors thank M. Hamar for his help with the experiment.

\end{document}